\documentclass[aps,prb,reprint,superscriptaddress,%
amsmath,amssymb,amsfonts]{revtex4-2}
\usepackage{array,multirow,graphicx,color,dcolumn,bm,hyperref}
\begin{document}
\title{Enhancement of spin-mixing conductance by $s$-$d$ orbital hybridization in heavy metals}
\author{Adam B. Cahaya}\email{adam@sci.ui.ac.id}
\affiliation{Department of Physics, Faculty of Mathematics and Natural Sciences, Universitas Indonesia, Depok 16424, Indonesia}
\author{Rico M. Sitorus}
\affiliation{Department of Physics, Faculty of Mathematics and Natural Sciences, Universitas Indonesia, Depok 16424, Indonesia}
\author{Anugrah Azhar}
\affiliation{Physics Study Program, Faculty of Sciences and Technology, Syarif Hidayatullah State Islamic University Jakarta, South Tangerang 15412, Indonesia}
\author{Ahmad~R.~T.~Nugraha}
\affiliation{Research Center for Quantum Physics, National Research and 
  Innovation Agency (BRIN), South Tangerang 15314, Indonesia}
\author{Muhammad Aziz Majidi}
\affiliation{Department of Physics, Faculty of Mathematics and Natural Sciences, Universitas Indonesia, Depok 16424, Indonesia}
\begin{abstract}
In a magnetic multilayer, the spin transfer between localized magnetization dynamics and itinerant conduction spin arises from the interaction between a normal metal and an adjacent ferromagnetic layer. The spin-mixing conductance then governs the spin-transfer torques and spin pumping at the magnetic interface. Theoretical description of spin-mixing conductance at the magnetic interface often employs a single conduction-band model. However, there is orbital hybridization between conduction $s$ electron and localized $d$ electron of the heavy transition metal, in which the single conduction-band model is insufficient to describe the $s$-$d$ orbital hybridization. In this work, using the generalized Anderson model, we estimate the spin-mixing conductance that arises from the $s$-$d$ orbital hybridization. We find that the orbital hybridization increases the magnitude of the spin-mixing conductance.
\end{abstract}
\keywords{spin-mixing conductance, spin pumping, spin transfer torque,  orbital hybridization, orbital mixing, electron-electron interaction}
\maketitle
\date{\today}
\section{Introduction}
The technological potential of magnetic devices based on transition metals for spin-current manipulation has pushed research forward in the spintronics area \cite{PhysRevB.72.024426}. The basic structure of a magnetic device is a magnetic multilayer. In spin-based memory systems, the interaction between normal metal and ferromagnetic metal can cause the magnetization direction to change \cite{SpinCurrent}.

In a magnetic multilayer, magnetization dynamics can be induced by spin current via the spin-transfer torque effect \cite{PhysRevB.77.224419}. When the non magnetic layer has a finite spin accumulation $\boldsymbol{\mu}$, which represents the difference of the spin-dependent electrochemical potential, the magnetization near the ferromagnetic interface experiences a torque $\boldsymbol{\tau}$ due to spin transfer \cite{PhysRevB.103.094420} 
\begin{align}
\boldsymbol{\tau}= g_{\uparrow\downarrow}\textbf{m}\times\left(\textbf{m}\times \boldsymbol{\mu}\right) \label{Eq.stt},
\end{align}
where $ g_{\uparrow\downarrow}$ is spin-mixing conductance. 
Reciprocally, in spin pumping, the spin current can be induced by magnetization dynamic $\textbf{m}$ via the exchange interaction between magnetization and spin of the conduction electron \cite{PhysRevB.66.224403}. An adiabatic precession of the magnetization pumps a spin current from the ferromagnet to the nonmagnetic layer with a polarization\cite{PhysRevLett.88.117601,Cahaya2021,CAHAYA2022JMMM}
\begin{align}
\textbf{J}=g_{\uparrow\downarrow} \textbf{m}\times \dot{\textbf{m}} .
\end{align} 
Both spin transfer torque and spin pumping effects are governed by the same $g_{\uparrow\downarrow}$, which has a complex value with a comparably small imaginary term \cite{PhysRevB.76.104409}.

\newpage

Spin-mixing conductance was originally described using spin-dependent scattering theory \cite{PhysRevB.66.224403}. The basic theoretical models of spin-mixing conductance utilizes a non-interacting electron model for the nonmagnetic metal \cite{PhysRevB.68.224403,PhysRevB.96.144434}. While this is certainly appropriate for free-electron-like metals, it is less so for heavy transition metals \cite{PhysRevLett.111.176601}. To accommodate the localized symmetry of the $d$ electron, a linear response theory description of spin-mixing conductance has been developed \cite{PhysRevB.96.144434}. However, there are few theoretical studies exploring the spin-mixing conductance of heavy metals with interacting electron model \cite{PhysRevB.88.054423}. Therefore, a better understanding of the spin-mixing conductance of heavy metals is required. 

\begin{figure}[ht]
\centering
\includegraphics[width=\columnwidth]{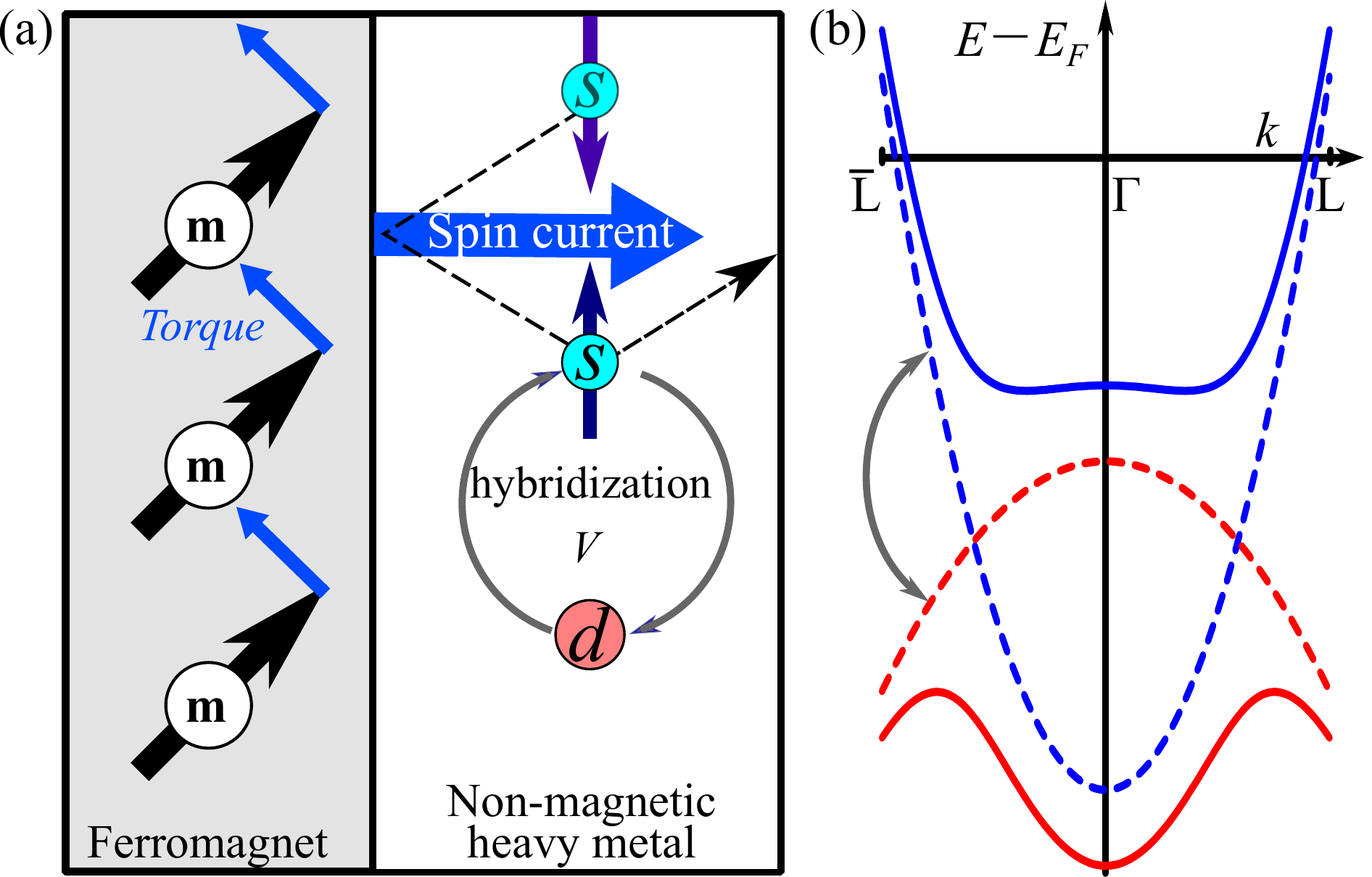}
\caption{(a) The interface of magnetic and heavy metals can be modeled as a ferromagnetic layer with localized magnetic moments and itinerant conduction electrons. The interaction of magnetic moments and conduction spin induces spin-transfer torque and spin current pumping in ferromagnetic layer and nonmagnetic layer, respectively. (b) In heavy metal, there is orbital hybridization between conduction $s$ electron and $d$ electron \cite{PhysRev.124.41} with parabolic dispersions\cite{PhysRev.120.67}. \label{Fig.Interface}}
\end{figure}

The theoretical description of spin-mixing conductance is often simplified in order to focus on a certain aspect or interaction that dominates a particular setup \cite{PhysRevLett.123.057203}. In the spin-pumping setup involving a heavy-metal system discussed in this article, we focus on the effect of electron-electron interaction at the nonmagnetic heavy metal layer. While the effect of electron interaction on spin-mixing conductance has been studied using Stoner model and phenomenological Hubbard parameter $U$ \cite{PhysRevB.50.7255,Zellermann_2004,Povzner2010}, a more realistic  model of the heavy-metal system requires orbital hybridization \cite{Sitorus_2021}, for example, the Anderson model \cite{PhysRev.124.41}. In the Anderson model, a $d$ electron is treated as an impurity, with well-localized energy dispersion \cite{PhysRev.149.491}. For describing a heavy metal, however, we need to consider a $d$ electron with  a more generalized dispersion \cite{PhysRev.120.67}. This article aims to theoretically estimate the electron-electron interaction correction factor due to the $s$-$d$ orbital hybridization of a heavy metal as illustrated in Fig.~\ref{Fig.Interface}.

In this article, we first analyze the linear response theory of spin density in heavy transition metal using Anderson model in Sec.~\ref{Sec.Susceptibility}. In Sec.~\ref{Sec.Spinmixing} we show that the orbital hybridization enhances $g_{\uparrow\downarrow}$ of the interface of ferromagnet and heavy transition metal (Ta, W, Ir, Pt or  Au). Lastly, we summarize our findings in Sec.~\ref{Sec.Conclusion}.

\section{linear response theory in heavy transition metal}
\label{Sec.Susceptibility}

Near the magnetic interface, the exchange interaction between the localized magnetic moments $\textbf{m}$ at the magnetic interface and the conduction spin $\boldsymbol{\sigma}$ can be written in the following Hamiltonian 
\begin{equation}
H_{ex}=-J\int d\mathbf{r}\delta(\mathbf{r})\textbf{m}\cdot\boldsymbol{\sigma}(\textbf{r}),
\end{equation}
where the exchange constant in the strong screening limit is $J=M_s/\left(\gamma \mathcal{N}_F\right)$  \cite{PhysRevB.96.144434,PhysRevB.103.094420}. $M_s$ and $\gamma$ are saturation magnetization and gyromagnetic ratio of the ferromagnetic insulator, respectively. $\mathcal{N}_F$ is the density of states of the conduction electron at the Fermi level. 

Using linear response theory, one can show the response of $\boldsymbol{\sigma}$ to $\textbf{m}$ via spin susceptibility $\chi_{ij}$
\begin{align}
\sigma_i(\textbf{r},t)=&J\int d\textbf{r}'dt'\chi_{ij}(\textbf{r}-\textbf{r}',t-t')m_j(t')\delta(\textbf{r}'), \label{Eq.sigma}
\end{align}
where 
\begin{align}
\chi_{ij}(\textbf{r},t)=\frac{i}{\hbar}\Theta(t)\left\langle \left[ \sigma_i(\textbf{r},t),\sigma_j(\textbf{0},0) \right]\right\rangle
\end{align}
Here $\Theta(t)$ is Heaviside step function. In an ,isotropic medium $\chi_{ij}=\delta_{ij}\chi$ can be written in terms of $\chi^{-+}$ and $\chi^{+-}$
\begin{align}
\chi=\chi^{-+}+\chi^{+-},
\end{align}
where
\begin{align*}
&\chi^{-+}(\textbf{r},t)=\frac{i}{\hbar}\Theta(t)\left\langle \left[ \sigma^-(\textbf{r},t),\sigma^+(\textbf{0},0) \right]\right\rangle,\\
&\chi^{+-}(\textbf{r},t)=\frac{i}{\hbar}\Theta(t)\left\langle \left[ \sigma^+(\textbf{r},t),\sigma^-(\textbf{0},0) \right]\right\rangle.
\end{align*}
Here $\sigma_{\pm}=(\sigma_x\pm i\sigma_y)/2$ and $\sigma_{a}$ ($a=x,y,z$) are Pauli matrices.
Since $\chi^{+-}$ can be obtained by replacing $+$ and $-$, it is convenient to discuss $\chi^{-+}$.

For a noninteracting simple metal with parabolic dispersion $E_\textbf{k}=E_0+\hbar^2k^2/2m^*$, $\chi$ in a small-$\omega$ limit is \cite{PhysRevB.96.144434}
\begin{align}
&\lim_{\omega\to 0}\chi_0(\textbf{q},\omega)
=\lim_{\omega\to 0}\sum_\textbf{k}\frac{f_{\textbf{k}}-f_{\textbf{k}+\textbf{q}}}{E_{\textbf{k}+\textbf{q}}-E_{\textbf{k}}+\hbar\omega+i0^+}\notag\\
&=\frac{mk_F}{\pi^2\hbar^3}\left(\frac{1}{2}+\frac{k_F^2-\left(\frac{q}{2}\right)^2}{2k_Fq}\ln\left|\frac{1+\frac{q}{2k_F}}{1-\frac{q}{2k_F}}\right|\right)
+i\omega \frac{m^2\Theta(2k_F-q)}{2\pi\hbar^3q}\notag\\
&\equiv \chi_0^r(q)+i\omega \chi_0^i(q), \label{Eq.realimagchi}
\end{align}
where $f_\textbf{k}$ is the low-temperature Fermi-Dirac distribution with wave vector $\textbf{k}$.

This single-band picture is appropriate for simple metals, such as light transition metals. Meanwhile, for heavy transition metal such as Au, W, Ta, and Pt, a localized $5d$-electron can mixed with the $6s$ band, as illustrated in the band structure [see Fig.~\ref{Fig.Interface}(b)]. The band structure can be obtained from density functional theory (DFT) software \cite{Castep,QE} (see the appendix).  Because of that, to determine the $g_{\uparrow\downarrow}$ of heavy metal, we need to modify the single-band Hamiltonian with an appropriate Hamiltonian that accommodates the hybridization of $s$ and $d$ electron.

In the second quantization, the interactions in a heavy-metal system near the interface that is illustrated in Fig.~\ref{Fig.Interface} can be written with the following Hamiltonian based on the Anderson model \cite{PhysRev.124.41,PhysRev.149.491} 
\begin{align}
H_0=&\sum_{\textbf{k}\alpha} \left(\begin{array}{cc}
a_{\textbf{k}\alpha}^\dagger & b_{\textbf{k}\alpha}^\dagger
\end{array}
\right)\left(\begin{array}{cc}
 E^s_{\textbf{k}}& V\\
V&  E^d_{\textbf{k}}
\end{array}
\right)\left(\begin{array}{cc}
a_{\textbf{k}\alpha}\\
b_{\textbf{k}\alpha}
\end{array}
\right)
 \label{Eq.Hamiltonian}
\end{align}
where $V$ is the hybridization parameter, $a_{\textbf{k}\alpha}^\dagger (a_{\textbf{k}\alpha})$ is the creation (annihilation) operator of $s$ electron with wave vector $\textbf{k}$ and spin $\alpha$ and $b_{j\alpha}^\dagger (b_{j\alpha})$ is the creation (annihilation) operator of the $d$ electron with spin $\alpha$. The second term corresponds to the $s$-$d$ hybridization. 
Here $\boldsymbol{\sigma}_{\alpha\beta}$ is Pauli vectors.  
The energy dispersion of $s$ and $d$ electrons can be assumed to be parabolic 
\begin{align}
    E^{s,d}_\textbf{k}=E^{s,d}_0+\frac{\hbar^2k^2}{2m^*_{s,d}},
\end{align}
as illustrated in Fig.~\ref{Fig.BandStructure}. 

\begin{table*}[ht]
\caption{Parameters of conduction electrons and its hybridization. The values are obtained using fitting from DFT \cite{Wannier90,QE,Castep} (see the appendix) \label{Table.parametermaterial}}
\begin{tabular}{cccccccccclcccc}
\hline
\hline\\[-2ex]
Heavy metal & crystal & \multirow{2}{*}{$a$ (\AA)} & \multirow{2}{*}{path} & \multirow{2}{*}{band} & \multicolumn{4}{c}{Without spin-orbit coupling} & & \multicolumn{4}{c}{With spin-orbit coupling} \\
\cline{6-9}\cline{11-14}
element   &    structure    &  &         &   & $V$ (eV) & $E^{s,d}_0$ (eV) & $m^*_{s,d}/m_e$ & $U_{sd}\mathcal{N}_F^s$ &   & $V$ (eV) & $E^{s,d}_0$ (eV) & $m^*_{s,d}/m_e$ & $U_{sd}\mathcal{N}_F^s$ \\[1ex]
\hline\\[-2ex]
\multirow{2}{*}{Au}	& \multirow{2}{*}{fcc}	& \multirow{2}{*}{2.88}	&  \multirow{2}{*}{$\Gamma-$L}& $s$ & \multirow{2}{*}{2.44}	& $-$8.86 & 1.19 & \multirow{2}{*}{0.09} 
& & \multirow{2}{*}{2.53}	& $-$8.23 & 1.28 & \multirow{2}{*}{0.15}   \\
  &	&       & 	& $d$ & 		                & $-$4.27 & $-$4.19	&      
 & &		& $-$4.05 & $-$5.46	        \\
\multirow{2}{*}{W }	& \multirow{2}{*}{bcc}& \multirow{2}{*}{2.74}	& \multirow{2}{*}{$\Gamma-$N}& $s$ & \multirow{2}{*}{2.38}	& $-$8.74 & 0.84 & \multirow{2}{*}{0.10}   
& & \multirow{2}{*}{2.34}		& $-$8.58 & 0.86 & \multirow{2}{*}{0.11}   \\
  	&       & &     & $d$ & 		                & $-$1.98 & $-$2.01 &        
& &		& $-$2.09 & $-$2.17        \\
\multirow{2}{*}{Ta}	& \multirow{2}{*}{bcc}& \multirow{2}{*}{2.87}	& \multirow{2}{*}{$\Gamma-$N}& $s$ & \multirow{2}{*}{2.28}	& $-$7.18 & 0.78 & \multirow{2}{*}{0.22}   
& & \multirow{2}{*}{2.25}	& $-$7.02 & 0.79 & \multirow{2}{*}{0.25}   \\
  	&       & &     & $d$ & 		                &  0.12 & $-$1.56 &
& &     &  -0.02 & $-$1.70         \\
\multirow{2}{*}{Ir}	& \multirow{2}{*}{fcc}& \multirow{2}{*}{2.74}	& \multirow{2}{*}{$\Gamma-$L}& $s$ & \multirow{2}{*}{3.76}	& $-$8.95 & 0.97 & \multirow{2}{*}{0.27}  
& & \multirow{2}{*}{3.62}	& $-$9.08 & 1.12 & \multirow{2}{*}{0.27}  \\
	&       & &     & $d$ &		                    & $-$2.94 & $-$2.96  &
& &     & $-$2.67 & $-$3.33         \\
\multirow{2}{*}{Pt}	& \multirow{2}{*}{fcc}& \multirow{2}{*}{2.77}	& \multirow{2}{*}{$\Gamma-$L}& $s$ & \multirow{2}{*}{3.22}	& $-$8.58 & 1.21 & \multirow{2}{*}{0.30}   
& & \multirow{2}{*}{3.32}	& $-$8.17 & 1.12 & \multirow{2}{*}{0.37}  \\
	&       &  &	& $d$ &		                    & $-$3.14 & $-$5.07  &       
& &     & $-$3.03 & $-$5.01        \\
\hline
\hline
\end{tabular}
\end{table*}

As illustrated in Fig.~\ref{Fig.Interface}, the $s$ electron dominates the spin-mixing process at the interface. Therefore, we can define $\boldsymbol{\sigma}(\textbf{r})$ from the spin density of the $s$ electron
\begin{align}
\boldsymbol{\sigma}(\textbf{r})
=&\sum_{\textbf{kq}\alpha\beta}e^{i\textbf{q}\cdot\textbf{r}}\boldsymbol{\sigma}_{\alpha\beta} a_{\textbf{k}+\textbf{q}\alpha}^\dagger a_{\textbf{k}\beta}.
\end{align} 
The susceptibility and its Fourier transform $\chi(\omega)=\int dt e^{i\omega t}\chi(t)$ can be determined by evaluating its time derivation using the Heisenberg equation
\begin{align}
\frac{\partial F(t)}{\partial t}=  \frac{1}{i\hbar}\left[F(t),H_0\right] \
\leftrightarrow \ \hbar\omega F(\omega)= \left[F(\omega),H_0\right].
\end{align}

Here $H_0$ is the unperturbed Hamiltonian in Eq.~(\ref{Eq.Hamiltonian}). Due to the hybridization of $s$ and $d$ orbitals, combinations of creation ($a^\dagger, b^\dagger$) and annihilation ($a, b$) operators appear when the commutations are evaluated. 
For convenience, we define
\begin{equation}
\chi_{abcd}^{-+}(\textbf{q},t)=\sum_\textbf{k}\chi_{abcd}^{-+}(\textbf{k},\textbf{q},t),
\end{equation} 
where \newpage
\begin{equation}
\chi_{abcd}^{-+}(\textbf{k},\textbf{q},t)=\frac{i}{\hbar}\Theta(t)\sum_{\textbf{k}'\textbf{q}'}\left\langle\left[a_{\textbf{k}+\textbf{q}\downarrow}(t)b_{\textbf{k}\uparrow}(t) , c_{\textbf{q}'\uparrow}(0)d_{\textbf{k}'\downarrow}(0) \right]\right\rangle.
\end{equation}
$\chi_{aaaa}$ in the frequency domain can now be obtained from a matrix relation
\begin{widetext}
\begin{align}
\left(\begin{array}{cccc}
E^s_{\textbf{k}+\textbf{q}}-E^s_{\textbf{k}}+\hbar\omega & $-$V & 0 & V \\
V & E^s_{\textbf{k}+\textbf{q}}-E^d_{\textbf{k}}+\hbar\omega & $-$V & 0 \\
0 & V & E^d_{\textbf{k}+\textbf{q}}-E^s_{\textbf{k}}+\hbar\omega & $-$V \\
-V & 0 & V & E^d_{\textbf{k}+\textbf{q}}-E^d_{\textbf{k}}+\hbar\omega\\
\end{array}
\right)
\left(\begin{array}{c}
\chi_{aaaa}\\
\chi_{abaa}\\
\chi_{baaa}\\
\chi_{bbaa}
\end{array}
\right)=
\left(\begin{array}{cccc}
f^{s}_{\textbf{k}}-f^{s}_{\textbf{k}+\textbf{q}}\\
0 \\
0 \\
0 
\end{array}
\right).
\end{align}
\end{widetext}
Let us note that since susceptibility is a retarded response \cite{CAHAYA2022JMMM}, $\omega$ has a negligibly small imaginary term $\omega=\lim_{\eta\to 0}(\omega+i\eta)$ as in Eq.~\ref{Eq.realimagchi}. By solving the linear equation, the following leading term of $V$-dependent spin susceptibility of the conduction $s$ electron $\chi_{aaaa}$ is
\begin{align}
\chi_{aaaa}(\textbf{k},\textbf{q},\omega)\simeq& \frac{\chi_0(\textbf{k},\textbf{q},\omega)}{1-U_{sd}(\textbf{k},\textbf{q})\chi_0(\textbf{k},\textbf{q},\omega)}\notag\\
\chi_0(\textbf{k},\textbf{q},\omega)=&\frac{f^s_{\textbf{k}}-f^s_{\textbf{k}+\textbf{q}}}{E^s_{\textbf{k}+\textbf{q}}-E^s_{\textbf{k}}+\hbar\omega+i0^+}.
\end{align} 
Here $U_{sd}$ is the electron-electron interaction parameter due to $s$-$d$ hybridization
\begin{align}
U_{sd}(\textbf{k},\textbf{q})=&\frac{\left(E^s_{\textbf{k}+\textbf{q}}+E^d_{\textbf{k}+\textbf{q}}-E^s_{\textbf{k}}-E^d_{\textbf{k}}\right)|V|^2}{\left(f^s_{\textbf{k}}-f^s_{\textbf{k}+\textbf{q}}\right)\left(E^s_{\textbf{k}+\textbf{q}}-E^d_{\textbf{k}}\right)\left(E^s_{\textbf{k}}-E^d_{\textbf{k}+\textbf{q}}\right)}.
\end{align}
The hybridization parameter $V$ can be obtained by fitting the band structure obtained from DFT using \textsc{wannier}90 software \cite{Wannier90}. The values of the parameters are listed in Table~\ref{Table.parametermaterial}. Since the spin-orbit interaction in a heavy metal is large, we also evaluate the parameters.

Using the localization of $\partial f_\textbf{q}/\partial E_\textbf{q}\approx-\delta(E-E_F)$, one can show that 
\begin{align}
\lim_{q,\omega\to 0}\chi(\textbf{q},\omega)&=\lim_{q\ll k}\sum_\textbf{k}\chi_{aaaa}^{-+}(\textbf{k},\textbf{q},\omega\to 0)\notag\\
&=\frac{\chi_0^r(0)}{1-U_{sd}\mathcal{N}_F^s}+\frac{i\omega\chi_0^i(q)}{\left(1-U_{sd}\mathcal{N}_F^s\right)^2},\label{Eq.Realchi}
\end{align}
where $\chi_0^r$ and $\chi_0^i$ are defined in Eq.~\ref{Eq.realimagchi} and
\begin{align}
U_{sd}\mathcal{N}_F^s=&\lim_{q\ll k_F}\sum_{\textbf{k}}U_{sd}(\textbf{k},\textbf{q})\chi_{aaaa}(\textbf{k},\textbf{q},0)\notag\\
=&\frac{\left(m^*_d+m^*_s\right)|V|^2}{m^*_d\left[E_s(k_F)-E_d(k_F)\right]^2} \label{Eq.Enhancement}
\end{align}
characterizes the enhancement due to the orbital hybridization.
This enhancement parameter is similar to the Stoner parameter $U\mathcal{N}_F$ that enhances the static magnetic susceptibility. 
Furthermore, for Au and W without spin-orbit interaction, \[U_{sd}\mathcal{N}_F^s\approx U\mathcal{N}_F,\] where $U$ is the phenomenological Hubbard parameter \cite{PhysRevB.50.7255} (see Table~\ref{Table.experimentcomparison}). However, $U_{sd}(k_F,0)\mathcal{N}_F^s<U\mathcal{N}_F$ for Ta, Ir and Pt. Table~\ref{Table.experimentcomparison} also shows that the spin-orbit interaction of the heavy metals increases $U_{sd}\mathcal{N}_F^s$.

\begin{table}[t]
\caption{spin-mixing conductance \cite{PhysRevLett.112.197201} and the enhancement factor due to orbital hyberidization of $5d$ heavy transition metals (HM) (see Table~\ref{Table.parametermaterial}). The electron-electron interaction parameter due to $s$-$d$ hybridization $U_{sd}\mathcal{N}_F^s$ (with and without SOI) is comparable to the Stoner parameter $U\mathcal{N}_F$ due to electron-phonon interaction \cite{PhysRevB.50.7255}.  \label{Table.experimentcomparison}}
\begin{tabular}{ccccccccccc}
\hline
\hline\\[-2ex]
HM & $U\mathcal{N}_F$\cite{PhysRevB.50.7255} & $U_{sd}\mathcal{N}_F^s$ & $g_{\uparrow\downarrow}\left(10^{18}\mathrm{m}^{-2}\right)$  & $g_{\uparrow\downarrow}\left(10^{19}\mathrm{m}^{-2}\right)$  \\
($5d$) &  & (with SOI) & Y$_3$Fe$_5$O$_{12}\vert$HM\cite{PhysRevLett.112.197201} & Co$\vert$HM\cite{PhysRevLett.120.157204}\\
\hline\\[-2ex]
Au	& 0.050	& 0.09 (0.15)	& 2.7$\pm$ 0.2 	&1.0$\pm$ 0.1 \\
W	& 0.102	& 0.10 (0.11)	& 4.5$\pm$ 0.4 	&1.2$\pm$ 0.1 \\
Ta	& 0.335	& 0.22 (0.25)	& 5.4$\pm$ 0.5 	&1.0$\pm$ 0.1 \\
Ir	& 0.290	& 0.27 (0.22)   &    -        	&2.4$\pm$ 0.2 \\
Pt	& 0.590	& 0.30 (0.37)	& 6.9$\pm$ 0.6 	&6.0$\pm$ 0.2 \\
\hline
\hline
\end{tabular}
\end{table}

\section{Enhancement of spin-mixing conductance}
\label{Sec.Spinmixing}

The spin current generation due to the exchange interaction between conduction spin and $\textbf{m}$ can be determined from the spin angular momentum loss due to the relative direction between conduction spin $\textbf{s}$ and $\textbf{m}$ \cite{Cahaya2021,CAHAYA2022JMMM}:
\begin{align}
\textbf{J}(t)=J\int d^3\textbf{r}\boldsymbol{\sigma}(\textbf{r},t)\times \textbf{m}(\textbf{r},t). \label{Eq.SxB}
\end{align}
Using the relation of $\chi$ and $\boldsymbol{\sigma}$, one can obtain the spin current from Eq.~(\ref{Eq.SxB}):
\begin{align}
\textbf{J}(t)=&J\int d^3\textbf{r}\boldsymbol{\sigma}(\textbf{r},t)\times \textbf{m}(\textbf{r},t)\notag\\
=& \textbf{m}(t)\times \dot{\textbf{m}}(t) \lim_{\omega\to0} J^2\sum_\textbf{qk} \frac{\partial\mathrm{Im}\chi_{aaaa}(\textbf{k},\textbf{q},\omega)}{\partial\omega}. \notag\\
\equiv& g_{\uparrow\downarrow} \textbf{m}(t)\times \dot{\textbf{m}}(t). 
\end{align}
Therefore, the spin-mixing conductance is enhanced by the orbital hybridization
\begin{align}
g_{\uparrow\downarrow}=& \lim_{\omega\to0} J^2\sum_\textbf{qk} \frac{\partial\mathrm{Im}\chi_{aaaa}(\textbf{k},\textbf{q},\omega)}{\partial\omega}
=\frac{J^2\sum_\textbf{q} \chi^i_0(q)}{\left(1-U_{sd}(k_F,0)\mathcal{N}_F^s\right)^2} .
\label{Eq.SpinMixingV}
\end{align}
Therefore the spin-mixing conductance is
\begin{align}
g_{\uparrow\downarrow}=& \frac{g_{\uparrow\downarrow}^0}{(1-U_{sd}\mathcal{N}_F^s)^2}.
\label{Eq.UsdN}
\end{align}
Here $U_{sd}\mathcal{N}_F^s$ is the effective electron-electron interaction parameter described in Eq.~(\ref{Eq.Enhancement}) and 
\begin{align}
g_{\uparrow\downarrow}^0\simeq\frac{\pi }{8}\left(\frac{M_s}{\gamma}\right)^2
\end{align}
is independent of the heavy metal \cite{PhysRevB.103.094420}. 

\begin{figure}[t]
\centering
\includegraphics[clip,width=0.9\columnwidth]{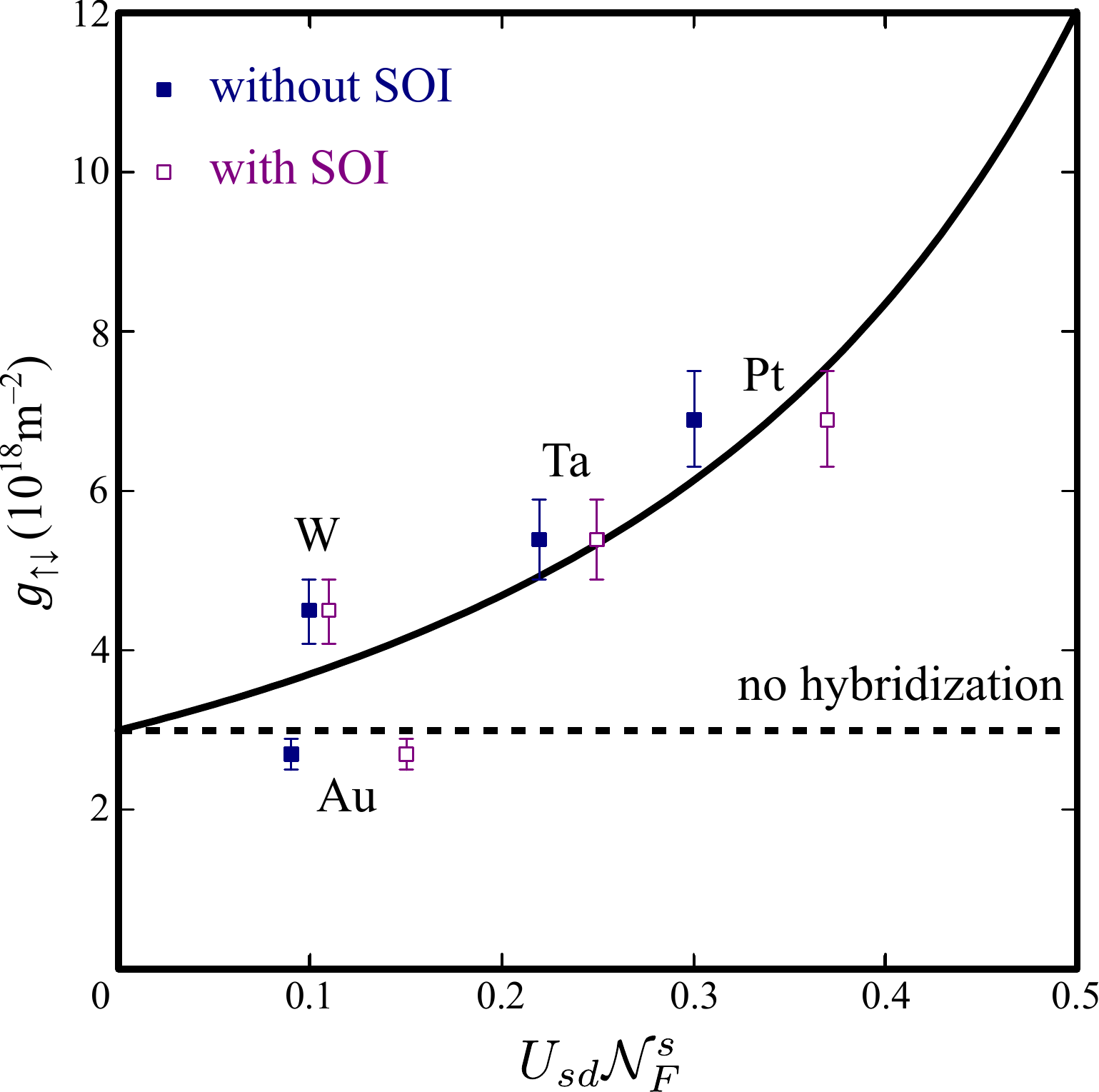}
\caption{Enhancement of spin-mixing conductance $g_{\uparrow\downarrow}$ of yttrium iron garnet (Y$_3$Fe$_5$O$_{12}$) and $5d$ heavy transition metal as a function $U_{sd}\mathcal{N}_F^s$, which characterizes the orbital hybridization. Filled square points are evaluated without spin-orbit interaction (SOI) while unfilled square points are values with SOI. Dashed and full lines are the values without and with hybridization, respectively.  Here $g_{\uparrow\downarrow}^0(\mathrm{Y_3Fe_5O_{12}})$ = 3 $\times10^{18}$ m$^{-2}$. The experimental data of the interface of Y$_3$Fe$_5$O$_{12}$ and $5d$ transition metals are taken from Ref.~\onlinecite{PhysRevLett.112.197201}, as summarized in Table~\ref{Table.experimentcomparison}. \label{Fig.SpinMixing}}
\end{figure}

\newpage
Figure~\ref{Fig.SpinMixing} shows the enhancement of spin-mixing conductance of an insulating ferromagnet Y$_3$Fe$_5$O$_{12}$ and a heavy metal (HM) as a function of $U_{sd}\mathcal{N}_F^s$. For Y$_3$Fe$_5$O$_{12}$ with the magnetic moment $M_{\rm Y_3Fe_5O_{12}}=3\mu_B$ and unit cell lattice constant $a_\mathrm{Y_3Fe_5O_{12}}= 5.4$ \AA \cite{Jain2013,Rodic1999}, spin-mixing conductance per unit area $g_{\uparrow\downarrow}^0/A$ can be estimated to be 
\begin{align}
\frac{g_{\uparrow\downarrow}^0\left(\mathrm{Y_3Fe_5O_{12}\vert HM}\right)}{A}=& ~\frac{\pi\left(M_\mathrm{Y_3Fe_5O_{12}}/\gamma_\mathrm{Y_3Fe_5O_{12}}\right)^2 }{8 a^2_\mathrm{Y_3Fe_5O_{12}}}\notag\\
\approx& ~3\times 10^{18} \mathrm{m}^{-2}.
\end{align}
The result is in agreement with the experimental work of Ref.~\onlinecite{PhysRevLett.112.197201}. This indicates that the $s$-$d$ orbital hybridization induces an effective electron-electron interaction on the conduction $s$ electron of the heavy transition metal and increases the spin-mixing conductance at its interface with a ferromagnetic insulator.

The discussion so far focuses on the case when the ferromagnetic layer is insulating. In an insulating magnetic interface, the orbital hybridization dominates the scattering for the interface of a ferromagnetic insulator and heavy metal, because only the heavy metal contributes to the conduction electrons. However, in the case of a metallic ferromagnet, the interactions of a conduction electron near the interface is more complicated. For a metallic ferromagnet (e.g., cobalt), to capture the complexity of the heavy-metal system~\cite{PhysRevLett.123.057203}, the enhancement factor should be replaced by a phenomenological parameter of the Stoner model\cite{PhysRevB.67.144418,PhysRevB.103.094420}. 
\begin{align}
g_{\uparrow\downarrow}\left(\mathrm{Co\vert HM}\right)=\frac{g_{\uparrow\downarrow}^0\left(\mathrm{Co\vert HM}\right)}{\left(1-U\mathcal{N}_F\right)^2}, \label{Eq.UN}
\end{align}
where
\begin{align}
\frac{g_{\uparrow\downarrow}^0\left(\mathrm{Co\vert HM}\right)}{A}=\frac{\pi d_{\rm Co}}{8  V_{\rm Co}}\left(\frac{M_\mathrm{Co}}{\gamma_\mathrm{Co} }\right)^2\approx 1.1\times 10^{19} \mathrm{m}^{-2}
\end{align}
is the unenhanced spin-mixing conductance of the bilayer of HM and Co with width $d_{\rm Co}=10$ \AA\cite{PhysRevLett.120.157204}, magnetic moment $M_{\rm Co}=1.60\mu_B$, and cell volume $V_{\rm Co}=22$ \AA$^3$ \cite{Jain2013,osti_1263614}.

\begin{figure}[t]
\centering
\includegraphics[width=0.9\columnwidth]{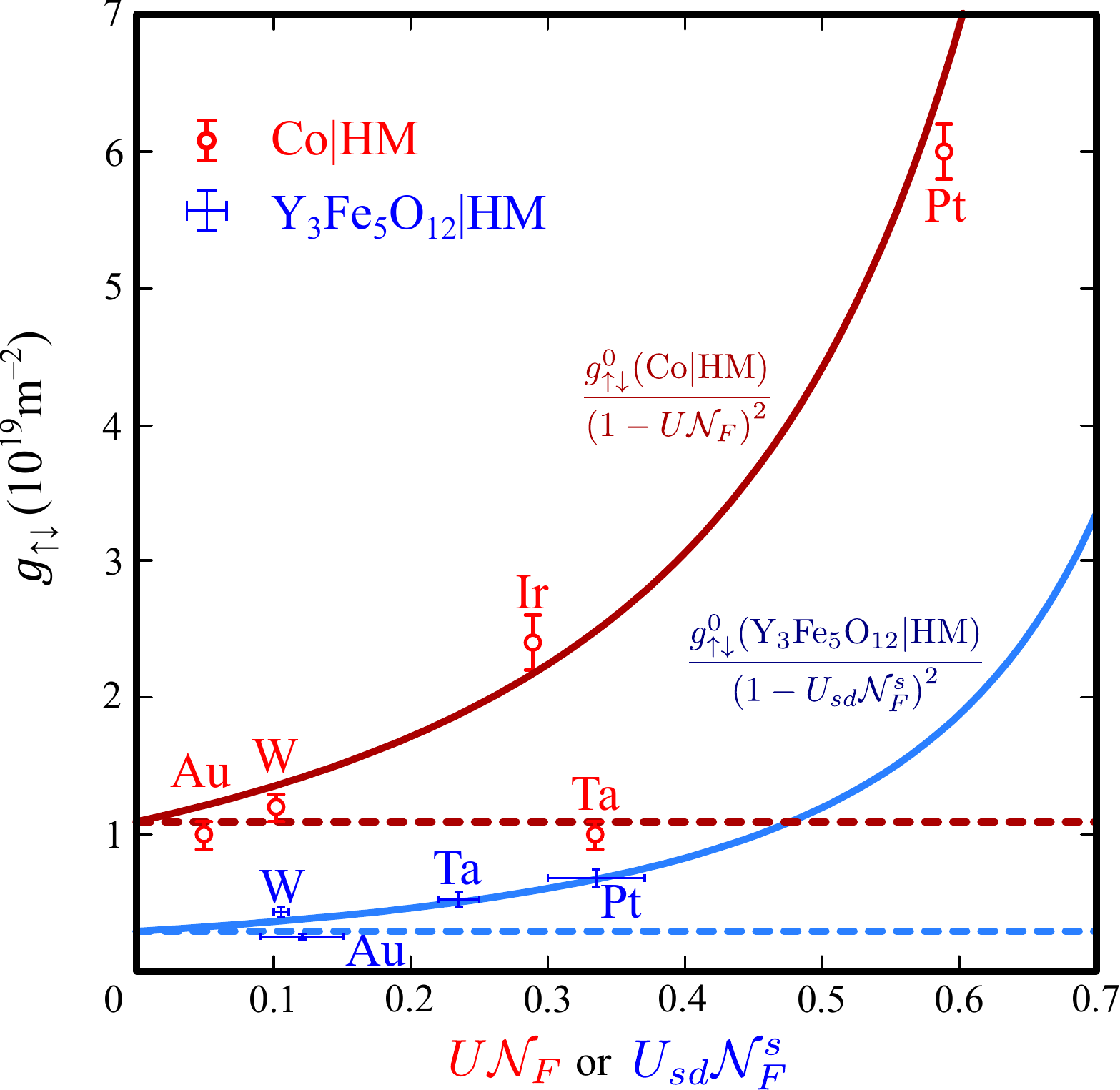}
\caption{Spin-mixing conductance $g_{\uparrow\downarrow}$ of (a) Y$_3$Fe$_5$O$_{12}\vert$heavy metal (HM) and (b) Co$\vert$HM. Dashed blue and red lines are the values without hybridization for  Y$_3$Fe$_5$O$_{12}$ and Co, respectively. Full blue and red lines show the values with hybridization for  Y$_3$Fe$_5$O$_{12}$ and Co, respectively. While Y$_3$Fe$_5$O$_{12}$ is an insulating ferromagnet, Co is a metallic ferromagnet. Experimental data of Y$_3$Fe$_5$O$_{12}\vert$HM and Co$\vert$HM are taken from Refs.~\onlinecite{PhysRevLett.112.197201} and \onlinecite{PhysRevLett.120.157204}, respectively (see Table~\ref{Table.experimentcomparison}). For a metallic ferromagnet such as Co, the enhancement is characterized by Stoner parameter $U\mathcal{N}_F$. On the other hand, for an insulating ferromagnet such as Y$_3$Fe$_5$O$_{12}$, the enhancement is dominated by $s$-$d$ hybridization $U_{sd}\mathcal{N}_F^s$ (averaged from values in Fig.~\ref{Fig.SpinMixing}).  \label{Fig.YIGandCo}}
\end{figure}

Figure~\ref{Fig.YIGandCo} shows the agreement of Eq.~\ref{Eq.UN} with the experiment of a metallic ferromagnet and Eq.~(\ref{Eq.UsdN}) with the experiment of insulating ferromagnet.
Co$\vert$HM has a larger $g_{\uparrow\downarrow}$ than Y$_3$Fe$_5$O$_{12}$ because the conduction spin can penetrate into a metallic ferromagnet and interact with more magnetic moments. When the ferromagnet layer is an insulator, the conduction electron purely originates from the heavy transition metal. Therefore, $s$-$d$ hybridization dominates the electron-electron interaction and our model is more appropriate.

\section{Conclusions}
\label{Sec.Conclusion}

To summarize, we discuss the effect of $s$-$d$ orbital hybridization on the spin-mixing conductance of the interface of ferromagnet and heavy metal. Using a generalized Anderson model, we study the linear response theory of conduction spin near a magnetic interface. At the magnetic interface, the hybridization of the conduction $s$ electron and localized $d$ electron of a heavy transition metal increases of the spin susceptibility of a heavy transition metal and subsequently enhances the spin-mixing conductance of the interface of ferromagnetic and $5d$ transition metal. 

For a bilayer of a ferromagnetic metal and a heavy metal, the enhancement of spin-mixing conductance is characterized by electron-electron interaction parameter in Stoner model $U\mathcal{N}_F$, as illustrated in Fig.~\ref{Fig.YIGandCo}. Meanwhile, for a bilayer of ferromagnet insulator and a heavy metal, the enhancement is characterized by the electron-electron interaction parameter $U_{sd}\mathcal{N}_F^s$ due to orbital hybridization that depends on the hybridization energy $V$ and the dispersion of $s$ and $d$ electrons. These parameters can be obtained by analyzing the band structure obtained from DFT. Figure~\ref{Fig.SpinMixing} shows the agreement of our theory and the experimental values of the bilayer of Y$_3$Fe$_5$O$_{12}$ and $5d$ transition metal. 

\begin{figure*}[t]
\centering
\includegraphics[width=0.85\textwidth]{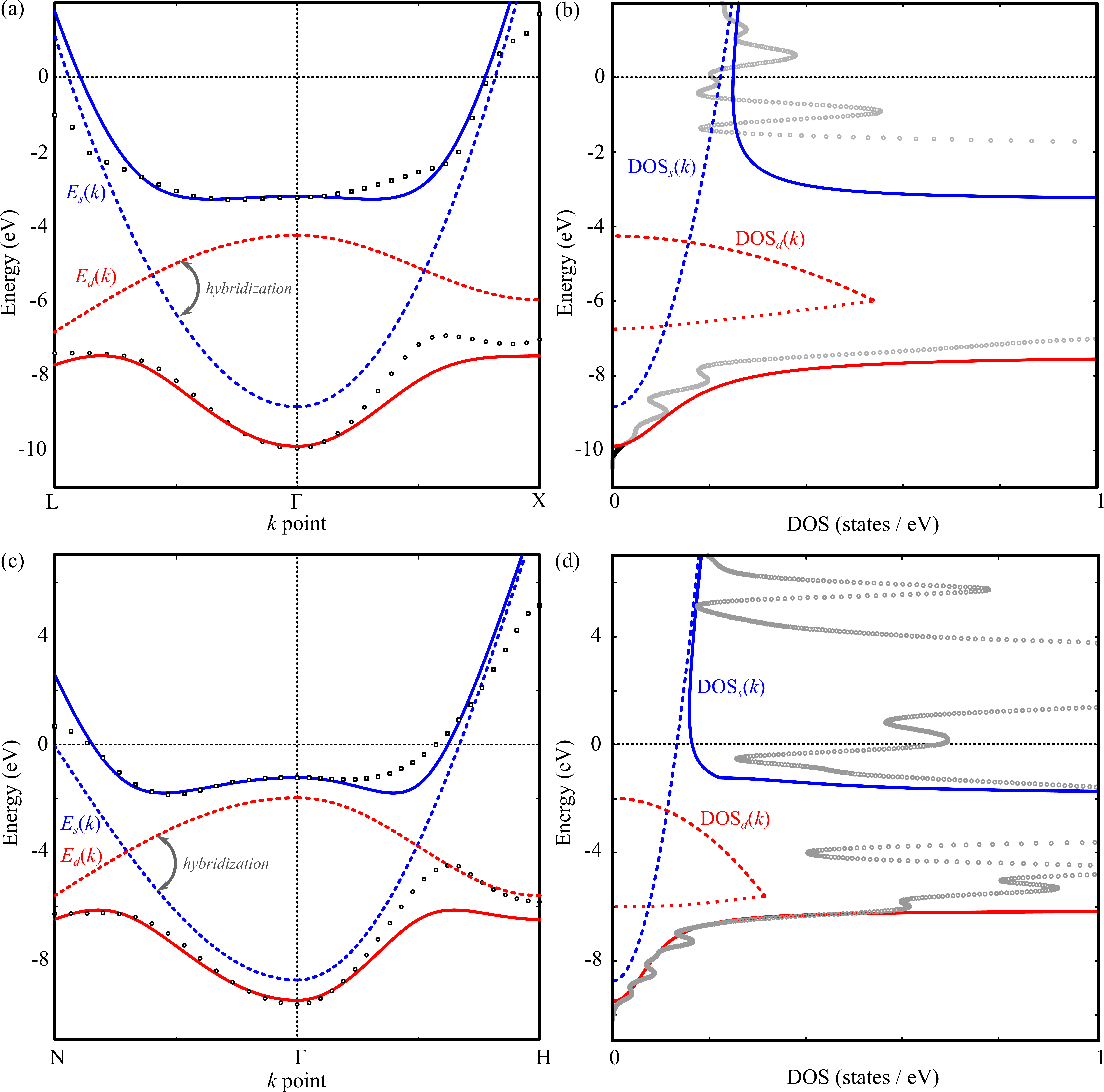}
\caption{Energy dispersion $E(k)$ and density of states (DOS) of Au and W. The energy dispersion and DOS of Au are shown in panels (a) and (b), respectively, while panels (c) and (d) show those of W. Data points were obtained using DFT. In panels (a) and (c), blue and red dotted lines indicate the energy dispersion of $s$ and $d$ electron without hybridization, respectively. The hybridized dispersion are illustrated with blue and red full lines. In panels (b) and (d), the dashed and full lines illustrate DOS without and with hybridization, respectively. While the energy dispersion only shows the hybridized band, the DOS shows the total DOS obtained using DFT. The orbital hybridization increases the DOS near the Fermi level. \label{Fig.BandStructure}}
\end{figure*}

\begin{acknowledgements}
We thank Universitas Indonesia for funding this research through PUTI Grant No. NKB-469/UN2.RST/HKP.05.00/2022.
\end{acknowledgements}
\appendix
\section*{\textsc{Appendix: Energy dispersion and density of states of 
\textit{5d} transition metals}}
\label{Sec:Appendix}

In this article we analyze the orbital mixing of Ta, W, Ir, Pt and Au. The orbital mixing occurs because of the hybridization between conduction ($s$ band) and localized ($d$ band, illustrated by DOS$_d$) electrons \cite{PhysRev.120.67}. The hybridized energy bands due to Hamiltonian in Eq.~(\ref{Eq.Hamiltonian}) are 
\begin{align}
    E_{12}(\textbf{k})=\frac{E^s_\textbf{k}+E^d_\textbf{k}}{2}\pm \sqrt{\left(\frac{E^s_\textbf{k}-E^d_\textbf{k}}{2}\right)^2+|V|^2}, \label{Eq.sdhybrid}
\end{align}
As illustrated in Fig.~\ref{Fig.BandStructure}, the partially filled band near the Fermi surface is chosen as $E_1(k)$, while the band at the bottom of the density of states is chosen as $E_2(k)$. Ir, Pt and Au have fcc structures. Figures~\ref{Fig.BandStructure}(a) and \ref{Fig.BandStructure}(b) illustrate the band structure along L$-\Gamma-$X symmetry points and density of states, respectively.  On the other hand, Ta and W have bcc structure. Figures~\ref{Fig.BandStructure}(c) and \ref{Fig.BandStructure}(d) illustrate the band structure along N$-\Gamma-$H symmetry points and density of states, respectively. 

By assuming $E^s_\textbf{k}$ and $E^d_\textbf{k}$ to be parabolic near $\Gamma$ point, the band structure parameters can be obtained by fitting the band structure obtained from DFT. 
The sum of $E_{1}$ and $E_{2}$
\begin{equation}
    E_{1}+E_{2}= E^s_\textbf{k}+E^d_\textbf{k} \equiv E^+_\Gamma+\frac{\hbar^2 k^2}{2m^*_+}
\end{equation}
can be used to obtain
\begin{align*}
    E^+_\Gamma=E_0^s+ E_0^d\\
    \frac{1}{m^*_+}=\frac{1}{m^*_s}+\frac{1}{m^*_d}.
\end{align*}
On the other hand, their difference
\begin{align}
    E_{1}-E_{2}=& \sqrt{\left(E^s_\textbf{k}-E^d_\textbf{k}\right)^2+4|V|^2} \notag\\
    \equiv& \sqrt{\left(E^-_\Gamma+  \frac{\hbar^2 k^2}{2m^*_-}\right)^2+4|V|^2} 
\end{align}
can be used to obtain 
\begin{align*}
E^-_\Gamma=E_0^s- E_0^d, \
\frac{1}{m^*_-}=\frac{1}{m^*_s}-\frac{1}{m^*_d},
\end{align*}
and hybridization energy $V$. The $E_0^{s'd}$ and effective masses can then be obtained from $E^\pm_\Gamma$ and $m^*_\pm$, respectively.


%

\end{document}